# Generative Adversarial Network (GAN) and Enhanced Root Mean Square Error (ERMSE): Deep Learning for Stock Price Movement Prediction

Ashish Kumar[1], Abeer Alsadoon[1,2,3,4*], P.W.C. Prasad[2,3], Salma Abdullah[5], Tarik A. Rashid[6], Duong Thu Hang Pham[7], Tran Quoc Vinh Nguyen[7]

[1]School of Computing and Mathematics, Charles Sturt University (CSU), Wagga Wagga, Australia
[2]School of Computer Data and Mathematical Sciences, University of Western Sydney (UWS), Sydney, Australia
[3]Kent Institute Australia, Sydney, Australia
[4]Asia Pacific International College (APIC), Sydney, Australia
[5]Department of Computer Engineering, University of Technology, Baghdad, Iraq
[6]Computer Science and Engineering, University of Kurdistan Hewler, Erbil, KRG, Iraq.
[7]The University of Da Nang – University of Science and Education, Faculty of Information Technology

Abeer Alsadoon[1*]
* Corresponding author. A/Prof (Dr) Abeer Alsadoon, [1]School of Computing and Mathematics, Charles Sturt University, Sydney, Australia.. Email: alsadoon.abeer@gmail.com , Phone +61 413971627

Abstract

The prediction of stock price movement direction is significant in financial circles and academic. Stock price contains complex, incomplete, and fuzzy information which makes it an extremely difficult task to predict its development trend. Predicting and analysing financial data is a nonlinear, time-dependent problem. With rapid development in machine learning and deep learning, this task can be performed more effectively by a purposely designed network. This paper aims to improve prediction accuracy and minimizing forecasting error loss through deep learning architecture by using Generative Adversarial Networks. It was proposed a generic model consisting of Phase-space Reconstruction (PSR) method for reconstructing price series and Generative Adversarial Network (GAN) which is a combination of two neural networks which are Long Short-Term Memory (LSTM) as Generative model and Convolutional Neural Network (CNN) as Discriminative model for adversarial training to forecast the stock market. LSTM will generate new instances based on historical basic indicators information and then CNN will estimate whether the data is predicted by LSTM or is real. It was found that the Generative Adversarial Network (GAN) has performed well on the enhanced root mean square error to LSTM, as it was 4.35% more accurate in predicting the direction and reduced processing time and RMSE by 78 secs and 0.029, respectively. This study provides a better result in the accuracy of the stock index. It seems that the proposed system concentrates on minimizing the root mean square error and processing time and improving the direction prediction accuracy, and provides a better result in the accuracy of the stock index.

# 1. Introduction

The stock market is an organized financial institute; shares and currency are bought and sold governed by the forces of demand and supply. The main purpose of a stock market is to serve as a primary and secondary market for an investor facilities price discovery and indicating economic activity. Stock price contains complex, incomplete and fuzzy information which makes it an extremely difficult task to predict its development trend [1]. Fluctuating financial data depend on a myriad of correlated constantly changing factors [2]. To predict the stock price, the traders use fundamental technical analysis methods. The fundamental analysis of exchange market suggests methods for the stock price to be used for prediction.

Several artificial neural network (ANN) models have been compared against the traditional model for stock exchange prediction [3]. The standard ANN technique was used in the last two decades in the stock exchange price prediction and trading strategies. However, various feature extraction methods were used for stock price prediction. Little researches have manifested that ANN has some limitations and it will not fit the stock price prediction ability. In another hand, the Deep learning techniques have addressed to overcome limitations and showed accurate results in several fields such as natural language, speech recognition and image analysis and recognition [3].

Deep Neural Network learning has been successfully implemented in several applications including speech and voice recognition and led towards to conclude that the deep learning techniques have performed well in the time-series data [4]. So, it seems possible that deep learning can be implemented to stock exchanges data as well. Nevertheless, accurate stock exchange prediction is considered chaotic. Furthermore, the logical analysis has shown that all current and new information results in the unpredicted stock price, thus this subject is still considered an open area for research [5]. So to attract an investor, an accurate predictive model is a major part of the stock exchange prediction model. Some requirements must be considered in defining an accurate model which is the direction prediction accuracy and forecast error during training and testing of the model. The studies of deep learning in the stock exchange have used a wide range of techniques and algorithms to improve the prediction accuracy and forecast error loss during the kerning and testing of the model.

The purpose of this paper is to enhance the prediction accuracy by minimizing the stock price direction prediction and forecast losses. The LSTM based RNN doesn't verify the correctness of the predicted value with the actual value during training and testing of the model leading to the increase in the value of Root Mean Square Error (RMSE). The research aims to improve the predictive performance of the stock price. This study proposes a Generative Adversarial Networks GAN as the combination of two neural networks (i.e. Generator and Discriminator) [6]. Stock data is typical time series, and the model G (Generator) is based on the LSTM model, which is broadly used for time series prediction. To estimate the probability of whether a sequence comes from the dataset or being predicted by generative model G, Convolutional Neural Network (CNN) was selected as discriminative model D to perform convolutional operations on one-dimensional input [6]. The main contributions of this study can be described as follows:

- Using the Loss function to measure the forecast error loss of each training sample.
- Directional Accuracy (DA) is the measurement of the accuracy relating to the series trend. It is used to measure the prediction capacity of the model. Higher the DA is higher the prediction accuracy. Or in other words, higher DA means more promising winning trades.
- MSRE is an indicator for the closeness of prediction to actual price.

The rest of the paper is organized as follows. Section 2 reviews the literature review. Section 3 proposes system and methods. Section 4 provides extensive experiments. Finally, give the conclusion in Section 5.

## 2. Literature Review

Yu and Yan [7] have compared four prediction methods which were deep LSTM model with no PSR, deep LSTM with PSR, deep multilayer perceptron (MLP) model conventional Support Vector Regression (SVR) machine learning method and the conventional ARIMA linear analytical method. The proposed method was a combination of PSR method with a DNN based LSTM model for time series analysis which improved the prediction accuracy. DNN based LSTM model showed prediction accuracy of an average of 56.85% which was 3-5% better than those of MLP and SVR models and much higher than that of ARIMA model. Although this model has shown higher prediction capacity than conventional machine learning algorithms. However, macroscopic factors and unstructured data were missing that need to be considered for far better accuracy and reliability. Other study [8] has extracted the hidden information from noisy text by initially normalizing it using Soundex phonetic algorithm with an inverse edit term frequency method. Then, it was analysed by character level Convolutional Neural Network (CNN) learning model. It has provided an accuracy of 75% in sentiment analysis and F-measure of 79%. This solution has helped to consider the impact of unstructured data (news, social media) and external factors while predicting the stock price. Unstructured data or factors have not been considered while making the stock price prediction model.

Zhou et al [6] have created a generic framework to predict the stock market by using Generative Adversarial Network (GAN) and used LSTM and CNN for adversarial training. Improved the stock price direction prediction accuracy to an average of 71.16% and minimize the forecast error (Root Mean Squared Relative error) to an average of 0.0079. Smaller model update cycles can improve the forecasting result by minimizing direction prediction loss. This work has showed a way to avoid complicated input preprocessing by adopting simple technical indexes. Previous study [3] has proposed design and architecture, based on market news and stock prices concurrently, of trading signal mining platform which implements Extreme Machine Learning (ELM). Kernelized ELM has achieved faster speed with a higher prediction of stock indexes than RBF SVM, BPNN, and basic ELM. This solution has focused on speed of prediction along with accuracy. Speed and accuracy are an important aspect of online trading. k-ELM has proved to have faster prediction speed and similar highest accuracy with RBF SVM.

Recent work [5] has elevated the accuracy and stability of stock price prediction. It has created a stock prediction model with 715 novel- input features and plunge filtering technique to improve accuracy. Accurate prediction of the stock price, by different techniques such as filtering techniques, target vector configuration and input- features used which were parts of the deep learning model. The model showed a high return rate of about 130% over a period of a year. However, the same model without the filtering technique has also showed high results. Therefore, plunge filtering technique did not make any high impact on returns. Moreover, several deep learning models such as CNN, RNN, GAN showed high return results were not considered. Göçken et al [9] have improved the prediction of highly complex stock price data. They have analyzed the stock market predictions by using Harmony Search for variable selection and parameter tuning then passing them to a hybrid model to hedge against potential risk in the market by testing harmony search with different prediction terms. Harmony Search – Jordan Recurrent Neural Network (HS-JRNN) outperforms the other prediction methods (Recurrent Extreme Learning Machine, Generalized Linear Model, Regression Tree, Gaussian Process Regression) and provides a promising direction to the study of stock market predictions. This solution has provided a way to select the number of neurons in hidden and context layers based on the input variables. As, not all parameters were considered such as training functions, iteration number which can highly affect the performance of the model.

Similar study [10] has offered a time-series information fusion framework, that is, an extended coupled hidden Markov model (ECHMM) for stock prediction. The solution has taken both stock events and price into consideration to improve the accuracy of prediction. It has provided an accuracy of 62.70% compared to 59.43% and 60.48% of ECHMM-NC (without using stock price information) and ECHMM-NE (without using stock event observation), respectively. This solution has included macroscopic factors as input for the prediction model, which has improved the result. However, sentiment factors have not been considered and they proved to be able to drive stock fluctuations. Other smilar work [11] has investigated the relationship between macroeconomic variables (inflation rate, money supply, exchange rate, interest rate) and the market return between Shanghai and Hong Kong by using Arbitrage Pricing Theory (APT), Vector Error Correction Model (VECM) and Granger-Causality test. It has tested the Johansen's cointegration test, VECM and Granger- causality and provided the outcomes as a long-term relationship, short term dynamics relationship, Granger-causality relationship, respectively. It implies that investors in the Chinese stock market should focus on long term investments whereas in Hong Kong stock market, the focus can be on both long-term and short-term.

Zhou et al [12] have enhanced the Support Vector Machine (SVM) model with based emotions selected (SVM-ES) to improve the prediction accuracy of the stock market. This solution has improved the accuracy of tradition SVM by integrating sentiment analysis. It revealed that inexperienced investors are 98% more sensitive to market fluctuations and SVM-ES provides an accuracy of 45.28% to 56.6% over traditional SVM. SVM-ES provides better results based on sentiment analysis. However, recent deep learning (CNN, RNN, GAN) models were not taken into consideration which tends to provide a more promising results than traditional methods. Other recent study [13] has introduced two deep CNN based frameworks, 2D-CNNpred and 3D-CNNpred, in order to improve prediction performance and accuracy by considering relation among different markets. This framework collects data from various sources, including different markets, for stock market prediction by extracting features. It has improves the prediction performance by about 3% to 11%, in terms of F-measure. However, it predicted the stock market, but no evidence provided for more fluctuating individual stock index.

Previous study [2] has proposed a novel methodology to improve the accuracy of market index forecast. It has conducted a research by forecasting market volatility using GARCH, and then Markov Switching combined with artificial neural fuzzy inference system (ANFIS) to determine the states of external factors and individual impact on each index, and ANN to improve the forecast accuracy. Suryoday et al [1] have developed a framework consisting of a random forest algorithm and gradient boosted decision trees to minimizing forecasting error and investment risk. This solution predicts the direction of stock price to signifying losses and gains. It was able to attain 78% accuracy. However, the proposed solution was only able to provide robust results for long term predictions. Derakhshan and Beigy [14] have improved Latent Dirichlet Allocation – part of speech (LDA-POS) by implementing part-of-speech into LDA to improve the accuracy of stock price prediction by assessing human sentiments in user reviews. It has provided an accuracy of 56.24% and 55.33% from English and Persian datasets, respectively. It has achieved better results than methods using explicit sentiment labels. However, other models have far better results than this model.

Sermpinis et al [15] have designed an adaptive SVR for stock data at different periods of time. The results found that the improved SVR with dynamic optimization of learning parameters by PSO can achieve higher prediction results than the traditional SVR. Moreover, machine learning models in the last years, are challenged by deep learning models in stock prediction [16].

## 2.1 State of the Art

Yu and Yan [7] have presented a system for predicting a stock index. Time series dataset usually contains insufficient data and the model is often over-fitted when trained using machine learning. To overcome this problem, it was used the common sliding window method for Phase-space Reconstruction (PSR). Prediction of the stock price was performed by deep learning framework with Long Short-Term Memory (LSTM) and Recurrent Neural Networks (RNN). Fig.1 illustrates "The blue borders show the good features of this state-of-the-art solution, and the red border refers to the limitation of it". This model consists of four major stages which were data-collection environment, pre-processing environment and prediction environment.

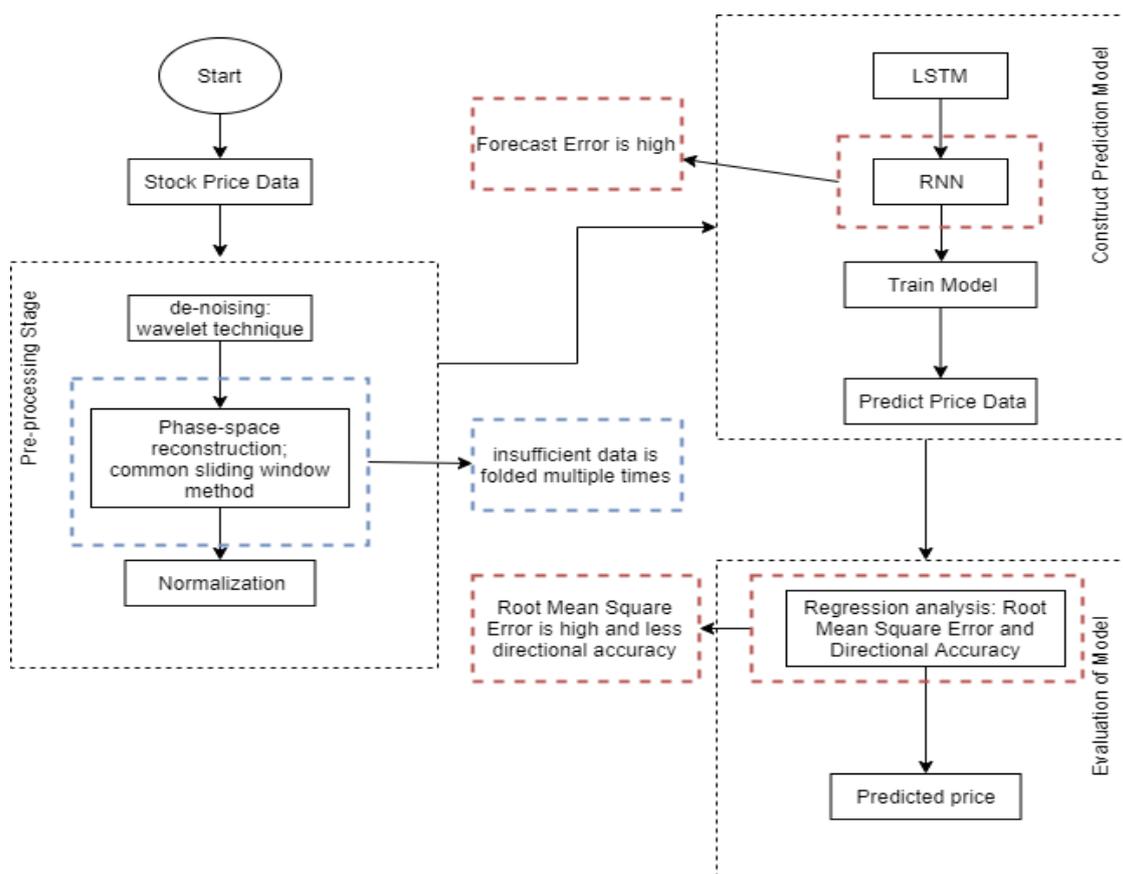

*Fig. 1: Block diagram of the state of art system: [7]*

*[The blue borders show the good features of this state-of-the-art solution, and the red border refers to the limitation of it]*

- *Data-Collection stage*: In this stage, the data is collected from Yahoo Finance (finanace.yahoo.com), TuShare financial data interface (tushare.org) and relevant organizations. Basically, the individual stock is considered during the prediction process because the company sharing data does not indicate the firm value.
- *Pre-Processing stage*: In this stage, data is de-noised by using wavelet technique. Time-series data is inconsistent, so PSR is used to fold and replicate the data multiple times using the common sliding window method. The dimensionality of price data is extended by PSR. The result is fed to the LSTM based Recurrent Neural Network (RNN) algorithm.
- *Prediction stage*: In this stage, two processes are mainly involved in any prediction system which are training and testing. LSTM is a basic deep learning model and capable of learning long term dependencies. An LSTM consists of so-called dynamic gate structures, an input gate, an output gate, and a forget gate [4]. RNN model using the LSTM node is effective and expendable when used to solve time series problems [7]. However, this model does not provide enough prediction accuracy because it fails to minimize the direction prediction loss and Root mean square error (RMSE) value. RMSE is frequently used to measure the difference between values predicted by LSTM based RNN and the observed values.
- *Evaluation Stage*: In this stage, the prediction result is evaluated on the defined metrics and presented in the form of charts. The prediction result is subject to multi-dimensional analysis.

The loss function is used to measure the forecast error loss of each training sample. This will only find the loss on the bases of the LSTM based RNN model. However, the number of iterations in this function increases the loss error and processing time.

$$L(Y(t+1), f(X(t)) = \frac{1}{m} \sum_{d=0}^{m} (x'(t+d\tau) - (x(t+d\tau))^2. \qquad (1)$$

Where:

$x'(t + d\tau)$ = last dimension of reconstruction vector

m = embedding dimension of generated series

t = time

$\tau$ = delay interval

n = time span that requires consideration

m = embedding dimension of generated series

$$DA = \frac{1}{n}\sum_{i=1}^{n} a_t, \text{when } a_t = \begin{cases} 1, & \text{if } (trend' = trend), \\ 0, & \text{otherwise.} \end{cases} \quad (2)$$

Where: trend' = predicted value

trend = actual value

the trend at time t is defined as:

$$trend_t = \begin{cases} 0, y_t - y_{t-1} < 0, \\ 1, y_t - y_{t-1} > 0, \\ trend_{t-1}, y_t - y_{t-1} = 0. \end{cases}$$

The Root Mean Square Error (RMSE) is the preferred method to measure the prediction power of a single stock. However, it is ineffective in performing a uniform comparison of multiple stocks.

$$RMSE = \sqrt{\frac{1}{n}\sum_{t=1}^{n}(y_t - y'_t)^2}, \quad (3)$$

Where:

RMSE = Root Mean Square Error

y' = predicted stock price

y = actual stock price

t = time

n = maximum lag of time

*Table 1: Pseudocode for state of art (LSTM algorithm)*

| |
|---|
| **Algorithm:** Long Short-Term Memory (LSTM) <br> **Input:** Normalized data <br> **Output:** Closed stock price |
| Begin <br> Step 1: Set input data weight: $W_i, W_f, W_c, W_o$ <br> Step 2: Set Recurrent data Weight: $R_i, R_f, R_c, R_o$ <br> Step 3: Set peephole weight: $V \in R^N$ <br> Step 4: Set Offset: $b_i, b_f, b_c, b_o \in R^N$ <br> Step 5: At time t, $x_t$ is the input and $y_t$ is the output of the node <br> Step 6: $f_t = \sigma(W_f x_t + R_f h_{t-1} + b_f)$ is the output of the forget gate at time t <br> Step 7: $i_t = \sigma(W_i x_t + R_i h_{t-1} + b_i)$ is the output of the input gate at time t <br> Step 8: $\acute{C}_t, C_t$ is the input and cell structure of the node at time t, respectively, which are expressed as: <br> $\acute{C} = \tanh(W_c x_t + R_c h_{t-1} + b_c)$ <br> $C = i_t \odot \acute{C}_t + f_t \odot c_{t-1}$ <br> Step 9: $O_t = \sigma(W_o x_t + R_o h_{t-1} + b_o)$ is the output of the output gate <br> Step 10: The final output $h_t$ of the node is expressed: $h = O_t \odot \tanh(C_t)$ <br> End |

# 3. Proposed System

This model consists of three major parts as shown in Fig.2 which are pre-processing, prediction and evaluation environments.. The blue borders show the good features of this state-of-the-art solution, and the green border refers to new proposed feature.

- *Data-Collection stage*: In this stage, the data is collected from Yahoo Finance (finanace.yahoo.com). Basically, the individual stock is considered during the prediction process because the company sharing data does not indicate the firm value.
- *Pre-Processing stage*: In this stage, data is de-noised by using wavelet technique. Time-series data is inconsistent, so Phase-space Reconstruction (PSR) is used to fold and replicate the data multiple times using the common sliding window method. The dimensionality of price data is extended by PSR. The result is fed to new proposed Generative Adversarial Network.
- *Prediction stage*: This stage consists of two main processes which are training and testing. Both processes are performed by using stock price dataset obtained in the data collection stage. Generative Adversarial Networks GAN is the combination of two neural networks (i.e. Generator and Discriminator) [6]. Stock data is typical time series, and the model G (Generator) is based on the LSTM model, which is broadly used for time series prediction. To estimate the probability whether a sequence comes from the dataset or being predicted by generative model G, Convolutional Neural Network (CNN) is selected as discriminative model (D) to perform convolutional operations on one-dimensional input.
- *Evaluation Stage*: In this stage, the prediction result is evaluated on the defined metrics and presented in the form of charts. The prediction result is subject to multi-dimensional analysis.

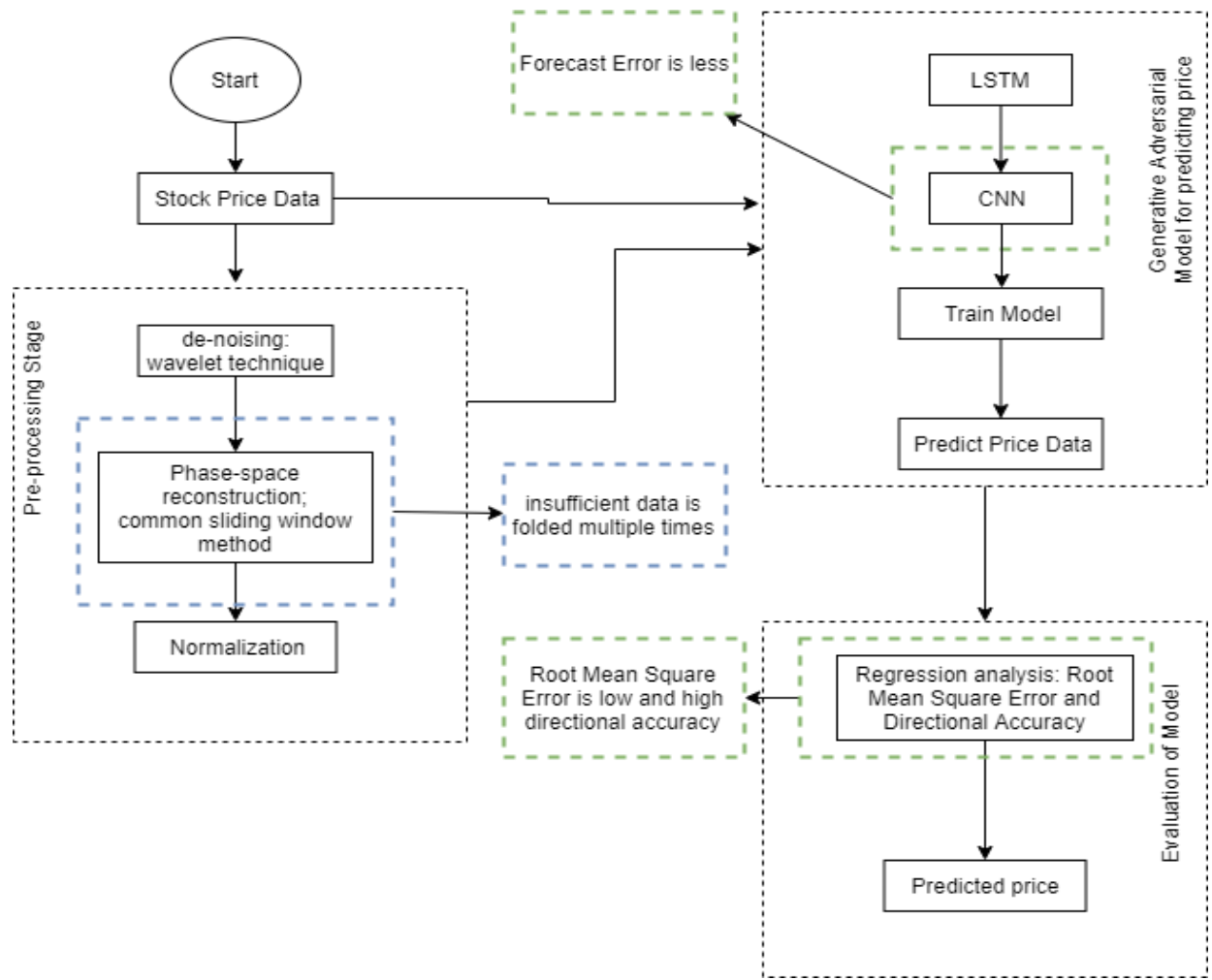

*Fig. 2: Block diagram of the new proposed system for stock price prediction using GAN*

*[The blue borders show the good features of this state-of-the-art solution, and the green border refers to new proposed feature]*

### 3.1. Proposed Equation

Loss Function expressed by [6] is the sum of all loss function in the model to minimize the forecast error loss of each training sampler. In order to make discriminative model D as confused as possible, adversarial loss function ($L_{adv}^G$) for generative model G should be minimized. Forecast Error Loss ($L_p$) should also be minimized to over the problem of being of confusing model D by model G without being close to $Y'_{T+1}$. $L_{dpl}$ is direction prediction loss function as stock price direction is very crucial for trading. There by combining all loss function, it was achieved final loss function (5).

$$L_G(X,Y) = \lambda_{adv} L_{adv}^G(Y') + \lambda_p L_p(Y,Y') + \lambda_{dpl} L_{dpl}(Y,Y'). \qquad (4)$$

Where:
$Y'$ = predicted output by generative model G
$Y$ = actual price
$L_{adv}^G$ is adversarial loss function for model G

$$L_{adv}^G(Y') = L_{sce}(D(Y'),1)$$

$L_{sce}$ is sigmoid cross-entropy loss function

$$L_{sce}(A,B) = -\sum_i B_i \log(sigmoid(A_i)) + (1-B_i)\log(1-sigmoid(A_i)),$$

$$L_{dpl}(Y,Y') = |sign(Y'_{T+1} - Y_T) - sign(Y_{T+1} - Y_T),$$

$$L_p(Y, Y') = ||Y - Y'||_p,$$

Where,
p = 1 or p = 2

Formula (5) is the modified version of formula (1)

$$ML(X,Y) = \lambda_{adv} L^G_{adv}(Y') + \lambda_p L_p(Y, Y') + \lambda_{dpl} L_{dpl}(Y, Y'). \tag{5}$$

Directional Accuracy (DA) is the measurement of the accuracy relating to the series trend. It is used to measure the prediction capacity of the model. Higher the DA is higher the prediction accuracy. Or in other words, higher DA means more promising winning trades.

$$DA = \frac{100}{T_0} \sum_{t=1}^{T_0} I_t, \text{ when } I_t = \begin{cases} 1, & \text{if } (Y_{t+1} - Y_t)(Y'_{t+1} - Y'_t) > 0, \\ 0, & \text{otherwise.} \end{cases} \tag{6}$$

where:
$T_0$ = total number of time points being tested
  $Y'$ = predicted stock price from generative model G
  Y = actual stock price

Below mentioned formula (7) is being modified result of equation (2)

$$MDA = \frac{100}{T_0} \sum_{t=1}^{T_0} a_t, \text{ when } a_t = \begin{cases} 1, & (Y_{t+1} - Y_t)(Y'_{t+1} - Y'_t) > 0, \\ 0, & \text{otherwise.} \end{cases} \tag{7}$$

where: $T_0$ = total number of time points being tested
  $Y'$ = predicted stock price from generative model G
  Y = actual stock price

Root Mean Square Relative Error (RMSRE) is an indicator of prediction accuracy. A low RMSRE which means that predicted data agrees with real. RMSRE facilitates the comparison of multiple stocks due to which this method is preferred over the RMSE.

$$RMSRE = \sqrt{\frac{1}{T_0} \sum_{t=1}^{T_0} \left(\frac{Y'_{t+1} - Y_{t+1}}{Y_{t+1}}\right)^2}, \tag{8}$$

where:
t = time
$T_0$ = total number of time points being tested
$Y'$ = predicted stock price from generative model G
Y = actual stock price

It was extracted formula from equation (8) and used it to enhance the equation (3). Formula (9) is the enhanced version. It calculated the loss function with the modification of root mean square error expression.

$$MRMSE = \sqrt{\frac{1}{T} \sum_{t=1}^{T} \left(\frac{Y'_{t+1} - Y_{t+1}}{Y_{t+1}}\right)^2}. \tag{9}$$

Where: MRMSE = Modified Root Mean Square Error
  t = time
  T = maximum lag of time
  Y' = predicted stock price from generative model G
  Y = actual stock price

### 3.2. Area of Improvement

In this study, it was considered to minimize the RMSE and forecast error loss which commonly occurs during the regulation in the calculation of the loss function by using equation 5. The algorithm has not performed well during the calculation of RMSE and forecast error loss due to the number of iterations (m) during training and testing of the model. However, several solutions were proposed to address this issue, but so far, more accurate minimization of forecast error loss and root mean square error (RMSE) has not been achieved.

Generative Adversarial Networks GAN is the combination of two neural networks (i.e. Generator and Discriminator) [6]. Stock data is typical time series, and the model G (Generator) is based on the LSTM model broadly used for time series prediction. To estimate the probability whether a sequence comes from the dataset or being predicted by generative model G, Convolutional Neural Network (CNN) was selected as discriminative model D to perform convolutional operations on one-dimensional input. The reason behind using adversarial loss is that it can mimic the behaviour of a financial trader. The available indicator is the main factor for an experienced trader for prediction of stock price then finds the true probability of his prediction with the old stock price, which is the work of generative model G and discriminative model D, respectively.

Table 2: Pseudocode the proposed model (GAN)

| |
|---|
| **Algorithm:** Proposed Generative Adversarial Network<br>**Input:** Normalized data<br>**Output:** Closed stock price |
| Begin<br>Step 1: Set learning rates $\rho_D$ and $\rho_G$, and parameters $\lambda_{adv}$, $\lambda_p$, $\lambda_{dpl}$;<br>Step 2: Initialize weights $W_D$ and $W_G$<br>Step 3: while not converged do<br>Step 4:     Update the Generator G;<br>Step 5:     Get K new data samples $(X^{(1)}, Y^{(1)}), (X^{(2)}, Y^{(2)}), …., (X^{(K)}, Y^{(K)})$<br>Step 6:     $W_G = W_G - \rho_G \sum_i^K \frac{\partial LG\ (X(1),Y(1))}{\partial WG}$<br>Step 7:     Update the discriminator D:<br>Step 8:     Get K new data samples $(X^{(1)}, Y^{(1)}), (X^{(2)}, Y^{(2)}), …., (X^{(K)}, Y^{(K)})$<br>Step 9:     $W_G = W_G - \rho_G \sum_i^K \frac{\partial LG\ (X(1),Y(1))}{\partial WG}$<br>Step 10: end while<br>End |

# 4. Result
## 4.1 Dataset

Python 3.7 IDE was used in the implementation and representation of the model by stock index dataset of various companies from 1 Jan 2013 to 31 Dec 2018. The dataset that taken as samples for testing has a different date (open, low, high, close, and volume data). Comprehensive examples which include different training epochs and batch size were undertaken. The dataset was obtained from Yahoo finance (finance.yahoo.com). Additionally the dataset considered an open-source and freely available on the internet as a (CSV file). The data will read it from the CSV file and fed it to a deep neural network. All the dataset has been tested in different window size as shown in fig 6, 7, 8, 9, 10. For the experiment, 23.2 GHz AMD Ryzen 5 2400GE with Radeon Vega Graphics and 8 GB RAM memory is used.

During the pre-processing phase, the wavelet technique has reduced the noise from dataset, folded volume ,data and replicated multiple times using common sliding window method to normalize the data. During the prediction phase, the Generative adversarial network was constructed by loading the dataset over the network. The generative model G has performed prediction $Y'_{T+1}$ on the input split dataset X. The predicted output $Y'_{T+1}$ and stock index closed price from CSV file were loaded to discriminative model D to perform convolution operations on the 1-D input sequence to estimate the probability whether time series comes from the dataset Y or being produced by a generative model G.

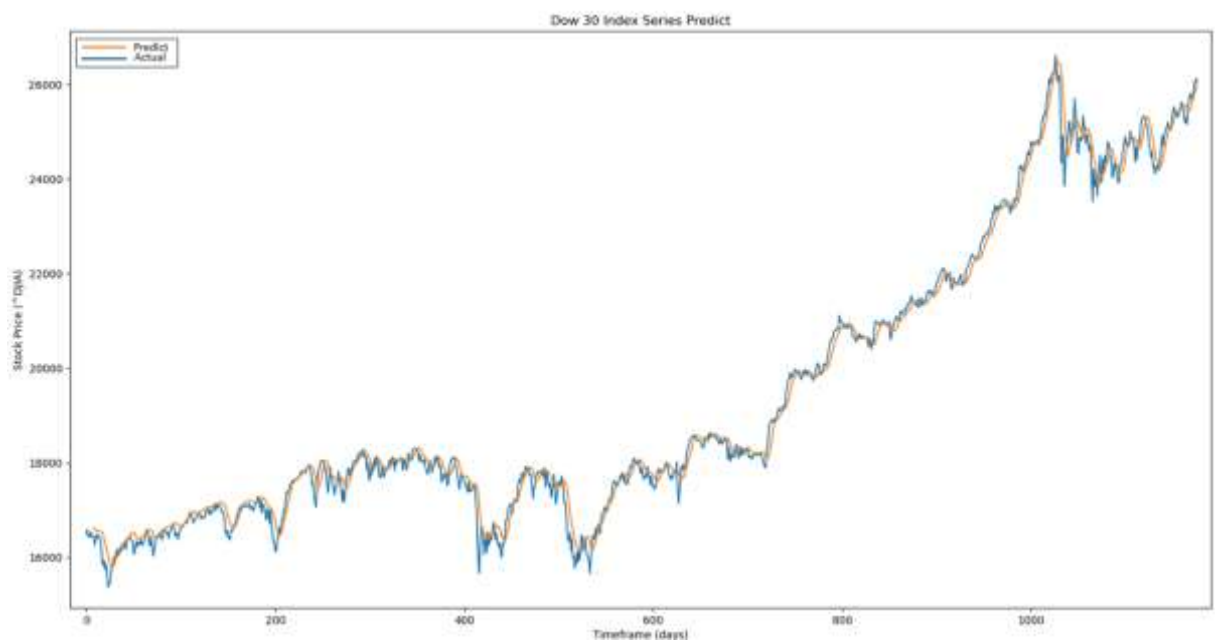

**Figure 6: Dow 30 Index Stock Predict**.

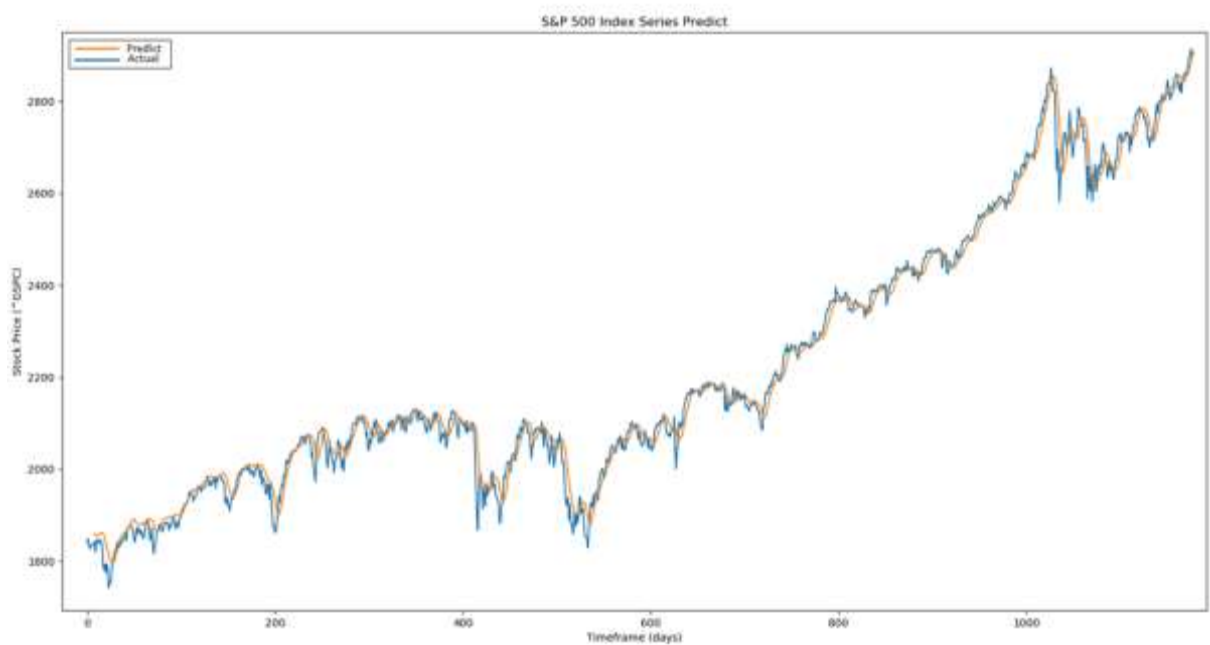

**Figure 7: S&P 500 Index Stock Predict**

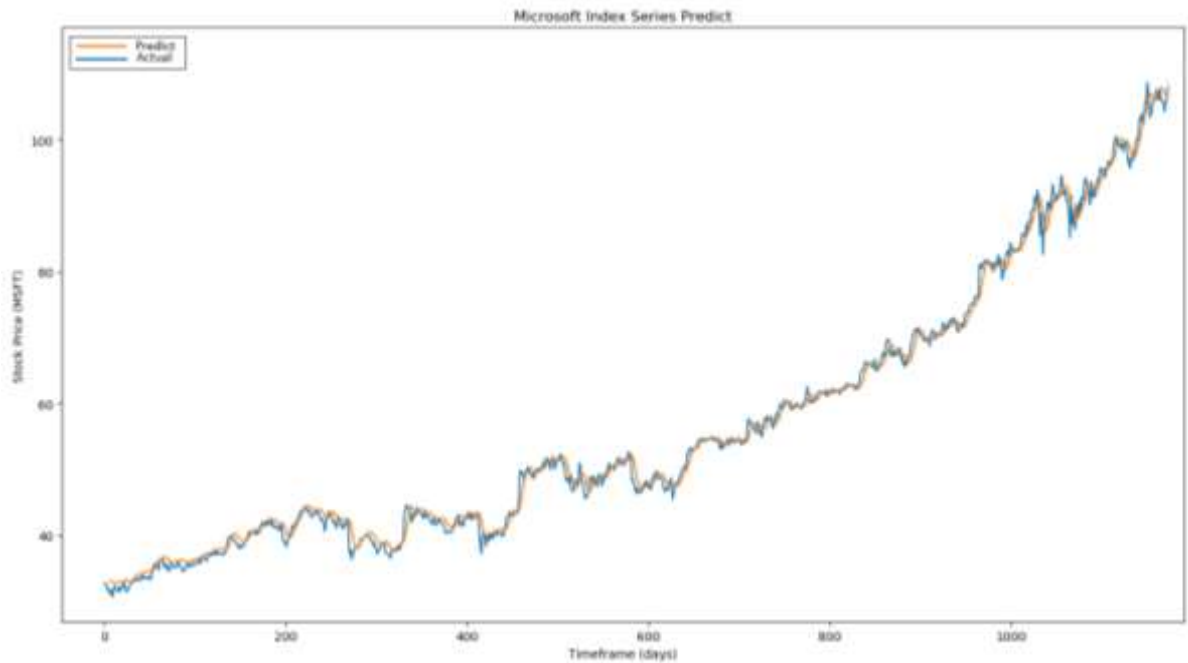

**Figure 8: MSFT Index Stock Predict**

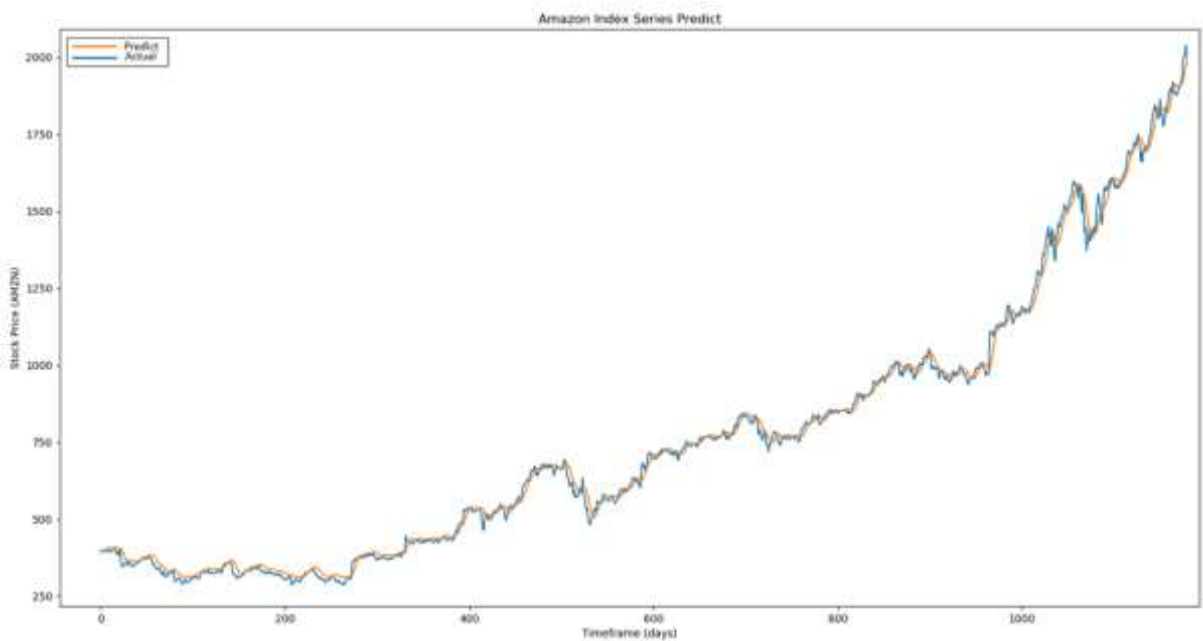

**Figure 9: Amazon Index Stock Predict**

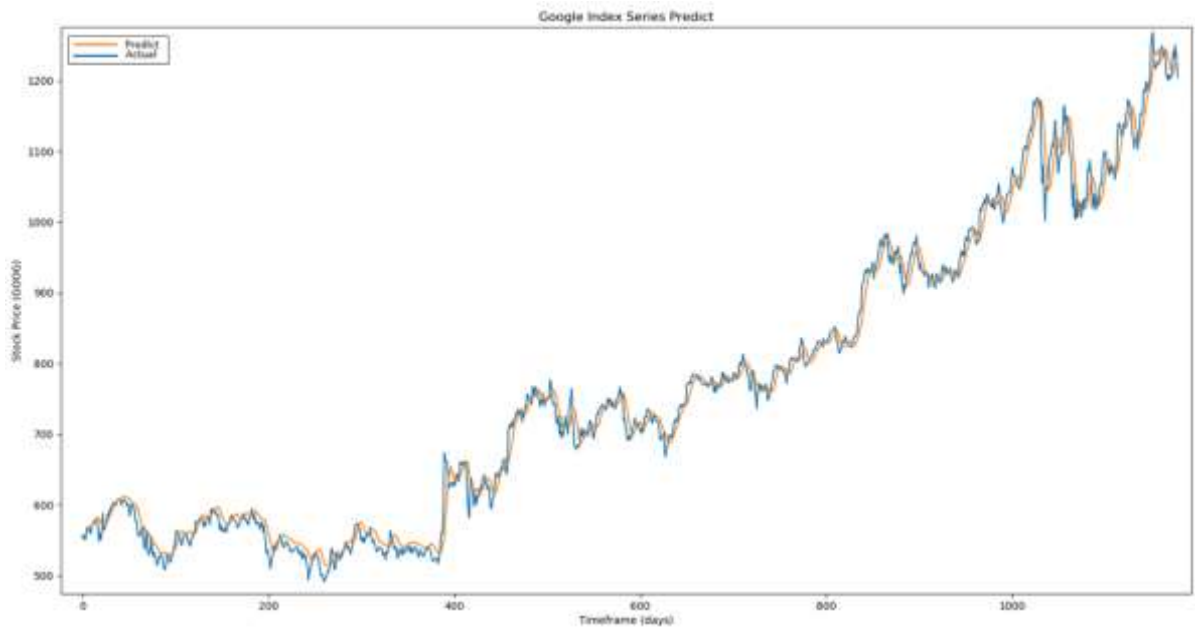

**Figure 10: Google Index Stock Predict**

### 4.2 Sub Result

The results of the current best state of the art solution and proposed solution were illustrated via flowcharts and tables. The results illustrated in figures 3, 4 and 5 shows the differences between the proposed solution and the best current solution. The result from the different stock exchange was reviewed in table 1 and figures 6, 7, 8, 9 and 10. All samples on the table contain the result obtained during prediction. This result was divided according to the training and testing purpose in the neural network. Here, the results from the sample were presented in terms of accuracy and processing time. Accuracy was determined in terms of root mean square error (RMSE), i.e. the difference between the actual target value and the predicted target value and the processing time was the number of iterations occurred during training and testing the model. A comprehensive test was done for 5 samples. The accuracy result was calculated by taking the average outcome of each test case in table 3. Then final result calculated by taking the average for all test cases of different scenarios. Proposal uses Generative Adversarial Networks and LSTM. According to their results, this new modification improves the model in terms of direction prediction and RMSE.

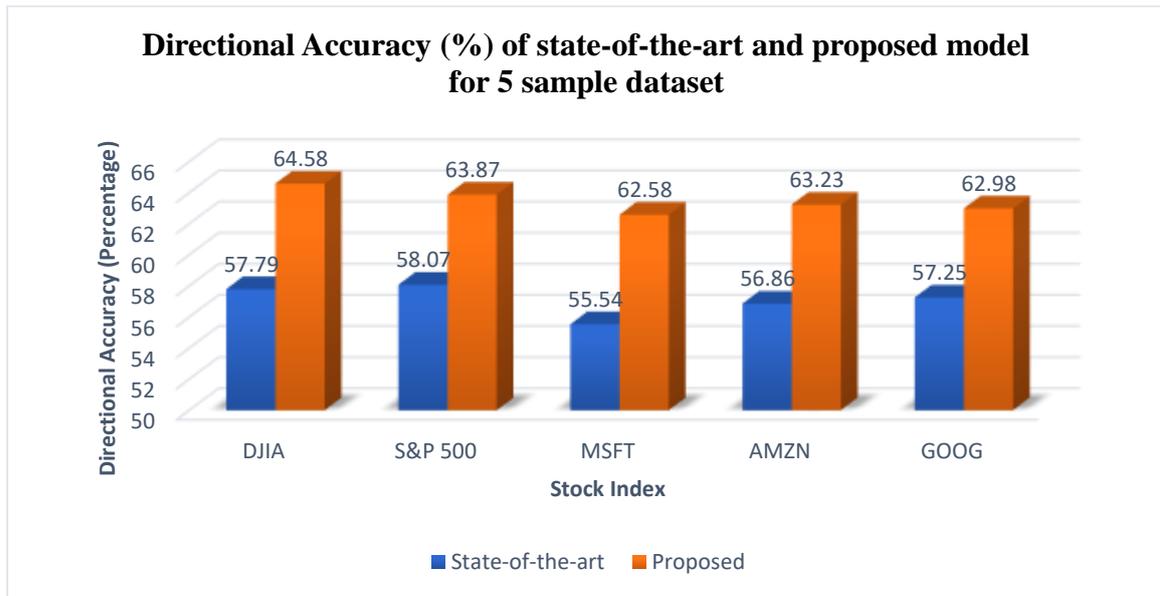

Fig 3: The directional accuracy in percentage for 5 sample stock index dataset using current LSTM model (blue) versus the proposed GAN-based model (red)

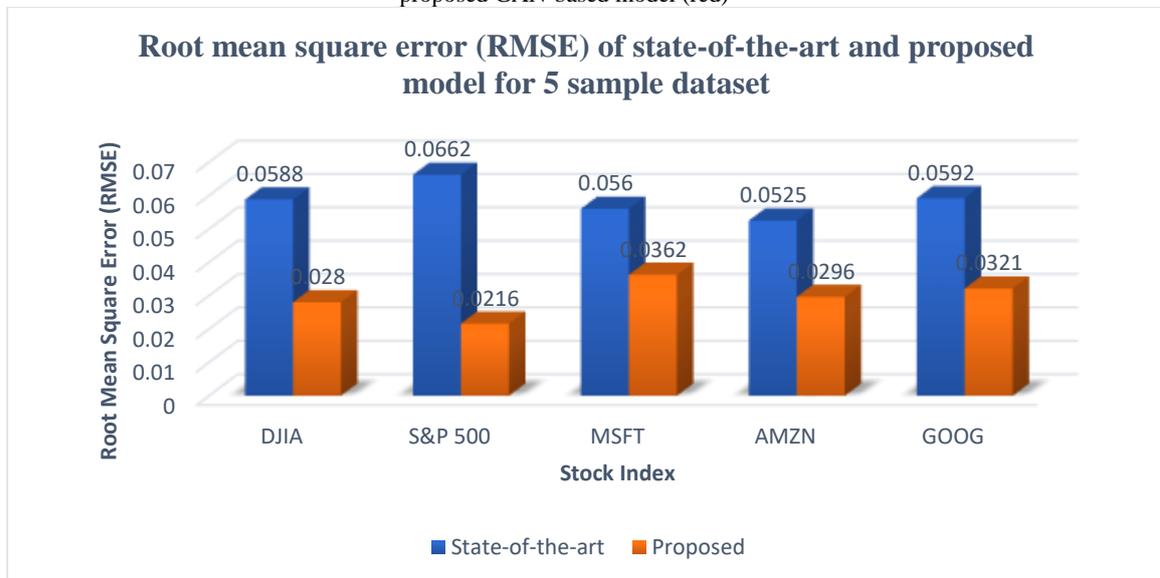

Fig 4: The root mean square error (RMSE) for 5 sample stock index dataset using current LSTM model (blue) versus the proposed GAN-based model (red).

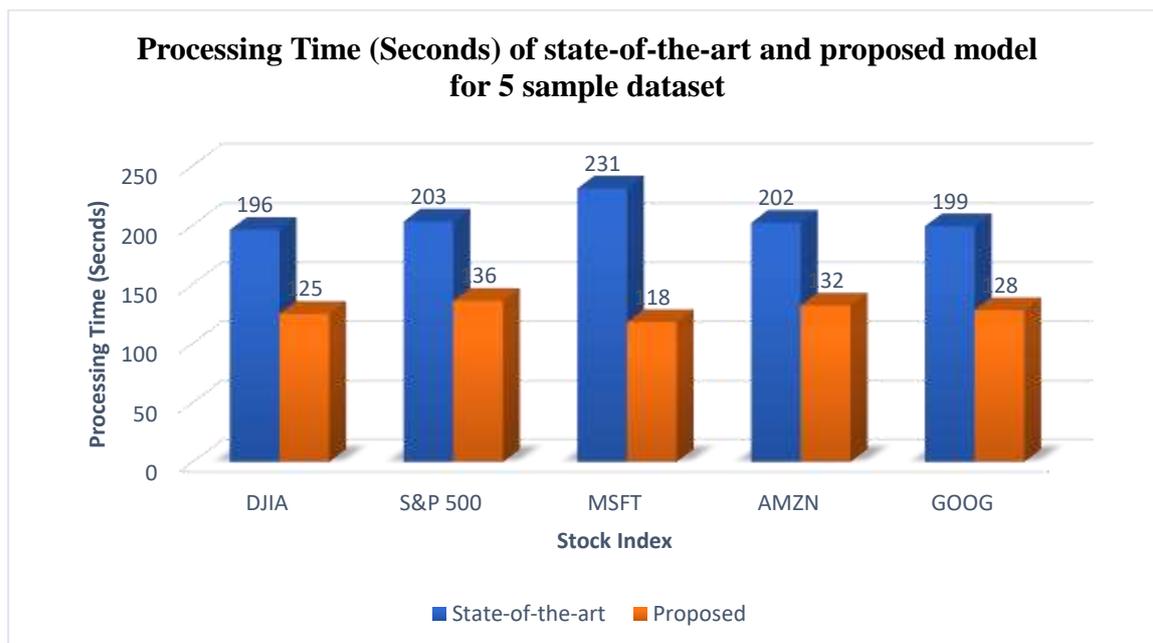

Fig 5: The processing time in seconds for 5 sample stock index dataset using current LSTM model (blue) versus the proposed GAN-based model (red)

**Table 1: Results of the comparison experiment on Index Stock Prediction**

| Stock Index | State of Art | | | Proposed | | |
|---|---|---|---|---|---|---|
| | Directional Accuracy | Processing Time | MRSE | Directional Accuracy | Processing Time | MRSE |
| DJIA | 57.79% | 196 Sec | 0.0588 | 64.58% | 125 Sec | 0.0280 |
| S&P500 | 58.07% | 203 Sec | 0.0662 | 63.87% | 136 Sec | 0.0216 |
| MSFT | 55.54% | 231 Sec | 0.0560 | 62.58% | 118 Sec | 0.0362 |
| AMZN | 56.86% | 202 Sec | 0.0525 | 63.23% | 132 Sec | 0.0296 |
| GOOG | 57.25% | 199 Sec | 0.0592 | 62.98% | 128 Sec | 0.0321 |

5. Discussion

Results show the difference in accuracy and processing time between the current state-of-the-art and the proposed solution concerning the accuracy and the processing time. The proposed algorithm has minimized the root mean square error value to 0.0295 from 0.0585. Besides, the average processing time was reduced from 3 min 26 sec to 2 min 8 sec due to reducing number of iterations. During the experimental tests, it was found that the LSTM has not performed well when it compared to the GAN. Although, GAN have performed very well, as the predicted value was close to the actual value. For the prediction of the stock index closed price, a range of techniques have been implemented, but continuously refined by the desire for accuracy and lower processing time.

This research has successfully overcome the limitations of the current best solution with 2 min and 8 sec against 3 min and 26 sec of average processing time. This research also has increased the directional accuracy to 61.45% against the current directional accuracy of 57.1%. This process was done by Phase-space reconstruction technique (common sliding window) for normalization of the data and fast processing. Additionally, the new feature has minimized the RMSE value and increased the directional accuracy of stock price prediction. The proposed system has been implemented and tested on Python programming language (Python 3.7) and shown to reduce the processing time and to increase accuracy. Along with this, the proposed solution with the actual and predicted result about the closed price helps to identify the closed stock price of following day. The above results show the differences between the accuracy and the processing time between the current solution and the proposed solution in terms of stock index closed price.

The system has been tested on Python 3.7 programming language, and the result shows that the proposed algorithm has successfully minimized the RMSE and forecast error loss. Also, it has reduced number of iterations during the training and the testing of the model which ultimately increases the processing time during the learning phase of the model. Based on the analysis results the GAN has used a window size of 121 and 100 epochs, as well as, it has used 25 hidden layers and 1 output. The values that have achieved by the RMSE are 0.0295 and the training period was 2 min 8 sec.

The comparative results of the state of the art and proposed solutions are shown in Table 2.

*Table 2: Comparative result of state-of-the-art and proposed result*

|  | **State-of-the-Art Solution** | **Proposed Solution** |
|---|---|---|
| **Applied Area** | Time series data | Time series data |
| **Features** | Reduce the loss function and tanh function | Reduce the loss function with modified root mean square error and linear rectification (ReLU) function |
| **Algorithm** | The algorithm is considered as a supervised learning | The algorithm is considered as an unsupervised learning |
| **Prediction Stage** | Use the output from Phase-space reconstruction algorithm and number of iterations to calculate the root mean square error | Use the output from Phase-space reconstruction algorithm and reduces the number of iterations to calculate the enhanced root mean square error (RMSE) |
| **Equation** | It calculated the loss function using equation-1 | It calculated the loss function with the modification of root mean square error expression from equation-9 |
| **Processing time** | The average processing time is 3 min and 26 sec | The average processing time is 2 min and 8 sec secs |
| **Accuracy** | The average directional accuracy is 57.10% and RMSE is 0.0585 | The average directional accuracy of the proposed model is 61.45% and RMSE is 0.0295 |

## 6. Conclusion

This paper presented combinations of Phase-space reconstruction (PSR) and GAN on the different form of dataset has improved the accuracy of stock prediction then it compares to the long short-term memory (LSTM) method with PSR.

The proposed method has been tested with diverse window size and epoch to find out the best parameters which can result in improving the system accuracy.

The experimental results show that the proposed GAN algorithm and deep learning perform well in the stock index closed price prediction. Although a range of techniques is available to predict the stock index close price, but they have so far failed to provide enough accuracy and processing time. Sufficient accuracy and low processing time considered major factors that affect the stock market prediction value.

This research has explored opportunities to overcome the limitations of the current best solution to achieve prediction accuracy based on RMSE equal to 0.0585 with average processing time equal to 3 min and 26 sec. This result justifies that the addition of the proposed ERMSE has overcome the limitations of the current best solution. Based on literature review, several solutions were proposed to address and overcome this issue, but so far none is able to provide an accurate root mean square error. The proposed algorithm has demonstrated capabilities to minimized RMSE to improve the directional prediction accuracy of the stock index closed price to 61.45% and 2 min 8 sec of processing time.

*Abbreviations*

| GAN | Generative Adversarial Network |
| CNN | Convolution Neural Network |
| RNN | Recurrent Neural Network |
| LSTM | Long Short-Term Memory |
| RMSE | Root Mean Square Error |
| 1D | One Dimensional |
| PSR | Phase-space Reconstruction |
| D | discriminative model |

# References

[1] B. Suryoday, S. Kar, S. Saha, L. Khaidem and S. R. Dey, "Predicting the direction of stock market prices using tree-based classifiers," *The North American Journal of Economics and Finance,* vol. 47, pp. 552-567, 2019.

[2] K. R. Werner and M. V. Kevin, "A stock market risk forecasting model through integration of switching regime, ANFIS and GARCH techniques," *Applied Soft Computing,* vol. 67, pp. 106-116, 2018.

[3] X. Li, H. Xie, R. Wang, Y. Cai, J. Cao, F. Wang, H. Min and X. Deng, "Empirical analysis: stock market prediction via extreme learning machine," *Neural Computing & Applications,* vol. 27, no. 1, pp. 67-78, 2016.

[4] W. Long, Z. Lu and L. Cui, "Deep learning-based feature engineering for stock price movement prediction," *Knowledge-Based Systems,* vol. 164, pp. 163-173, 15 January 2019.

[5] S. Yoojeong, L. Jae and L. Jongwoo, "A study on novel filtering and relationship between input-features and target-vectors in a deep learning model for stock price prediction," *Applied Intelligence,* vol. 49, no. 3, pp. 897-911, 2019.

[6] X. Zhou, Z. Pan, G. Hu, S. Tang and C. Zhao, "Stock market prediction on high-frequency data using generative adversarial nets," *Mathematical Problems in Engineering,* vol. 2018, pp. 1-11, 2018.

[7] P. Yu and X. Yan, "Stock price prediction based on deep neural networks," *Neural Computing and Applications,* pp. 1-20, 2019.

[8] M. Arora and V. Kansal, "Character level embedding with deep convolutional neural network for text normalization of unstructured data for Twitter sentiment analysis," *Social Network Analysis and Mining,* vol. 9, no. 1, pp. 1-14, 2019.

[9]  M. Göçken, M. Özçalıcı, A. Boru and A. T. Dosdoğru, "Stock price prediction using hybrid soft computing models incorporating parameter tuning and input variable selection," *Neural Computing and Applications,* vol. 31, no. 2, pp. 577-592, 2019.

[10] X. Zhang, Y. Li, S. Wang, B. Fang and P. Yu, "Enhancing stock market prediction with extended coupled hidden Markov model over multi-sourced data," *Knowledge and Information Systems,* pp. 1-10, 2018.

[11] Y. Ning, L. C. Wah and L. Erdan, "Stock price prediction based on error correction model and Granger casuality test," *Cluster Computing,* pp. 1-10, 2018.

[12] Z. Zhou, K. Xu and Z. Jichang, "Tales of Emotion and stock in China: volatility, causality and prediction," *World Wide Web,* vol. 21, no. 4, pp. 1093-1116, 2018.

[13] E. Hoseinzade and S. Haratizadeh, "CNNpred: CNN-based stock market prediction using a diverse set of variables," *Expert Systems with Applications,* vol. 129, pp. 273-285, 2019.

[14] A. Derakhshan and H. Beigy, "Sentiment analysis on stock social media for stock price movement prediction," *Engineering Applications of Artificial Intelligence,* vol. 85, pp. 569-578, 2019.

[15] G. Sermpinis, A. Karathanasopolulos, R. Rosillo and D. d. l. Fuente, "Neural networks in financial trading," *Annals of Operations Reasearch,* pp. 1-16, 2019.

[16] S. Jeon, B. Hong and V. Chang, "Pattern graph tracking-based stock price prediction using big data," *Future Generation Computer Systems,* vol. 80, pp. 171-187, 2018.